\documentclass[12pt]{article}

\usepackage{etex}
\usepackage{fullpage}
\usepackage{setspace}
\usepackage{comment}
\usepackage[pdftex]{graphicx}
\usepackage{rotating}
\usepackage{amsmath}
\usepackage{amsthm}
\usepackage{amssymb}
\usepackage{graphicx}
\usepackage{fullpage}
\usepackage{setspace}
\usepackage{natbib}
\usepackage{color}
\usepackage{booktabs}
\usepackage[linesnumbered, ruled]{algorithm2e}
\SetKwRepeat{Do}{do}{while}%

\usepackage{pgf}
\usepackage{pgfplots}
\usepackage{textpos}
\usepackage{graphicx}
\usepackage{pstricks}
\usepackage{framed}
\usepackage{varwidth}
\usepackage[normalem]{ulem}
\usepackage{xstring}

\newcommand{\T}{\intercal}

\newcommand{\OMIT}[1]{\relax}   
\def\text{{\rm}}

\def\cov{\hbox{Cov}}

\newcommand{\bma}[1]{\mbox{\boldmath $#1$}}

\newcommand{\bA}{ {\bma{A}} }
\newcommand{\ba}{ {\bma{a}} }

\newcommand{\bd}{ {\bma{d}} }

\newcommand{\bH}{ {\bma{H}} }
\newcommand{\bh}{ {\bma{h}} }

\newcommand{\bW}{ {\bma{W}} }

\newcommand{\bX}{ {\bma{X}} }
\newcommand{\bx}{ {\bma{x}} }

\theoremstyle{definition}

\def\cov{\hbox{Cov}}

\begin{document}

\title{Using Pilot Data to Size Observational Studies for the Estimation of Dynamic Treatment Regimes}


\pagestyle{empty}
\begin{center}
	\textbf{Using Pilot Data to Size Observational Studies for the Estimation of Dynamic Treatment Regimes} \\
	\textbf{Eric J. Rose$^{1}$, Erica E. M. Moodie$^{1}$, 
		Susan Shortreed$^{2,3}$}
	\\
	$^1$Department of Epidemiology and Biostatistics, McGill University,
	Montreal, QC, H3A 1A2, Canada  \\
	$^2$Kaiser Permanente Washington Health Research Institute, Seattle, WA, 98101, U.S.A.\\
	$^3$Department of Biostatistics, University of Washington, Seattle, WA, 98195, U.S.A.
\end{center}

\begin{abstract} \noindent
There has been significant attention given to developing data-driven methods for tailoring patient care based on individual patient characteristics. Dynamic treatment regimes formalize this through a sequence of decision rules that map patient information to a suggested treatment. The data for estimating and evaluating treatment regimes are ideally gathered through the use of Sequential Multiple Assignment Randomized Trials (SMARTs) though longitudinal observational studies are commonly used due to the potentially prohibitive costs of conducting a SMART. These studies are typically sized for simple comparisons of fixed treatment sequences or, in the case of observational studies, {\em a priori} sample size calculations are often not performed. We develop sample size procedures for the estimation of dynamic treatment regimes from observational studies. Our approach uses pilot data to ensure a study will have sufficient power for comparing the value of the optimal regime, i.e. the expected outcome if all patients in the population were treated by following the optimal regime, with a known comparison mean. Our approach also ensures the value of the estimated optimal treatment regime is within an {\em a priori} set range of the value of the true optimal regime with a high probability. We examine the performance of the proposed procedure with a simulation study and use it to size a study for reducing depressive symptoms using data from electronic health records. \\
\begin{footnotesize}
\end{footnotesize}
\end{abstract}

\pagebreak
\setcounter{page}{1}
\pagestyle{plain}
\section{Introduction}
\label{s:intro}

Data-driven methods for personalizing treatment assignment have been of great interest to clinicians and researchers. Dynamic treatment regimes (DTRs) operationalize clinical decision making through a sequence of decision rules that map up-to-date patient information to a recommended treatment \citep{dtr_book, Chakraborty_Moodie}. An optimal treatment regime is a set of decision rules that maximizes the mean of some measure of positive health outcome when all patients in the population of interest are assigned treatment by following that regime \citep{murphy2003optimal, robins2004optimal}. DTRs have been studied to improve decision making in healthcare across many different areas of applications such as for cancer \citep{zhao_2011, wang_2012}, schizophrenia \citep{shortreed_2012}, and depression \citep{chakraborty_2016}.

Many different methods for estimating optimal treatment regimes have been proposed, three of which are given by
Q-learning, G-estimation, and dynamic weighted ordinary least squares (dWOLS). Q-learning is a regression-based approach that is straightforward to implement in practice, but is not robust to model misspecification \citep{watkins1992q}. G-estimation uses a contrast function to estimate an optimal regime, and is double robust, meaning it is robust to misspecification of either the outcome or propensity score model, but is more complex to implement \citep{robins2004optimal}. dWOLS is based on a series of weighted regression models and is straightforward to implement in a continuous outcome setting, like Q-learning, while maintaining double-robustness properties \citep{Wallace_2015}.

While data for the estimation of optimal treatment regimes is ideally gathered through a Sequential Multiple Assignment Randomized Trial (SMART) \citep{lavori2000design, lavori2004dynamic, murphy2005experimental}, longitudinal observational studies are more commonly used because the resources for conducting SMARTs are often prohibitive and estimation of DTRs is often considered exploratory in nature \citep{chakraborty_2014}. Currently, SMARTs are typically sized for comparisons of specific treatment regimes under study, as opposed to identifying optimal regimes \citep{oetting_2011, lei2012smart, artman_2020, seewald_2020}. Observational studies are generally powered for the primary research question of the study. Estimation of an optimal DTR is not taken into account for sample size calculations for either type of study. This has resulted in many studies of DTRs failing to find a statistically significant benefit to tailoring treatments to individual patient characteristics \citep[e.g.][]{krakow_2017, simoneau_2020, coulombe_2020}. Further, there is no guarantee that the performance of an estimated regime will be close to that of the true optimal regime. Therefore, it has been advised that estimated optimal treatment regimes be evaluated with a follow-up study in which patients are randomized to regimes of interest \citep{murphy2005experimental}. This approach is costly since it requires conducting two studies and, if the original study is not powered to guarantee a high quality estimate of the optimal regime, the follow-up study could focus on a poor quality treatment regime.

There has been some work on sizing randomized trials for the estimation of optimal dynamic treatment regimes, i.e. estimating treatment strategies that are potentially more highly tailored than the simple, embedded strategies within the trial. These methods provide a sample size that ensures sufficient power for a comparison between the optimal regime and standard of care as well as ensuring that the performance of the estimated optimal regime is close to that of the true optimal regime. \cite{laber2016using} proposed a method for using pilot data to size a two-armed randomized trial that is based on inverting a projection confidence interval. \cite{Rose_SMART} proposed two methods for sizing two-stage SMARTs for the estimation of optimal dynamic treatment regimes. The first imposes strong assumptions on the underlying data generating model that assumes away the complexities related to non-regularity which leads to a sample size estimator that resembles a comparison of fixed treatment sequences. The second makes minimal assumptions and uses bootstrap oversampling (i.e. resampling with replacement with sample size greater than the original data size) with pilot data to estimate a sample size. Methods for sizing observational studies for the estimation of dynamic treatment regimes do not exist. 

In this paper, we propose a method for sizing a multi-stage longitudinal observational study for the estimation of dynamic treatment regimes using pilot data. This approach is based on constructing a projection interval and using bootstrap oversampling to find a sufficient sample size. This work provides the first sample size procedure for estimating an optimal treatment regime from observational data as well as the first procedure to size any study for the estimation of an optimal regime that consists of more than two treatment decisions. Our sample size estimation procedure requires a finite-sample (non-asymptotic) estimate of the variance of DTR parameters, which can be accomplished even in non-regular settings via the $m$-out-of-$n$ bootstrap \citep{Bibhas_2013}. However this approach has yet to be implemented in more than two stages. Thus, a further contribution of this work is to demonstrate, for the first time, this resampling method for estimating confidence intervals for the parameters indexing a dynamic treatment regime with more than two stages. 

We give an overview of the setup and notation for this work in Section \ref{s:setup}. In Section \ref{s: methodology}, we present the proposed sample size procedure. In Section \ref{s: simulation}, we provide a simulation study to demonstrate the empirical performance of the proposed approach and the variability in the resulting sample size across different pilot studies for a three-stage observational study. In Section \ref{s: KPWA}, we demonstrate the use of our method for sizing a study for reducing depressive symptoms using electronic health record (EHR) data from Kaiser Permanente Washington (KPWA). In Section \ref{s: discussion}, we give concluding remarks and a discussion of open problems.

\section{Setup and Notation}
\label{s:setup}

We consider sizing an observational study for estimating an optimal dynamic treatment regime with $K$ sequential treatment decisions. We assume two potential treatment options at each stage. The observed data are of the form $\mathcal{D}_n = \{(\bX_{1,i}, A_{1,i},\dots, \bX_{K,i}, A_{K,i}, \allowbreak Y_i )\}_{i=1}^{n} $ which is comprised of $n$ $i.i.d$
replicates of $(\bX_{1}, A_{1},\dots, \bX_{K}, \allowbreak A_{K}, Y)$ where: $\bX_1 \in \mathbb{R}^{p_1}$ denotes baseline patient information, $A_k \in \{0,1\}$ denotes the treatment assigned at the $k^{th}$ stage, $\bX_k \in \mathbb{R}^{p_k}$ for $k=2,\dots,K$ denotes additional patient information recorded during the course of treatment $(k - 1)$, and $Y \in \mathbb{R}$ denotes the outcome of interest, coded such that higher values are better. Let $\bH_k$  be patient history available to a clinical decision maker at stage $k$, so $\bH_1 = \bX_1$ and $\bH_k = (\bX_1^T, A_1, \dots, A_{k-1}, \bX_{k}^T)^T$ for $k=2,\dots,K$.

A treatment regime, $\bd$, is defined as a set of decision rules $\bd = (d_1, \dots, d_K)$, such that $d_k: \mathrm{dom}\,\bH_k \rightarrow \mathrm{dom}\, A_k $ for $k=1,\dots,K$ is a function that maps a patient's history to a recommended treatment lying within the domain (dom) of possible treatments. Therefore a patient with history $\bH_k = \bh_k$ is recommended to be assigned treatment $d_k(\bh_k)$. An optimal regime is one that maximizes the mean outcome when applied to the population of interest.

It is helpful to be able to reference a patient's treatments received and history up to or after a certain stage. To do this we use an overbar to denote the treatments, covariates, and regimes up to stage $k$ so we have that $\bar{\ba}_k = (a_1, \dots, a_k)$, $\bar{\bx}_k = (x_1, \dots, x_k)$, and $\bar{\bd}_k = (d_1, \dots, d_k)$. When we are considering the entire sequence of $K$ stages we will suppress the subscript so that $\bar{\ba} = \bar{\ba}_K$ and $\bar{\bx} = \bar{\bx}_K$. An underbar will denote treatments, covariates, and regimes from stage $k$ to $K$ such that $\underline{\ba}_k = (a_k, \dots, a_K)$, $\underline{\bx}_k = (x_k, \dots, x_K)$, and $\underline{\bd}_k = (d_k, \dots, d_K)$.

To formalize the notion of an optimal regime we use the potential outcome framework \citep{rubin_1978}. Let $\bH_k^*(\bar{\ba}_{k-1})$ be the potential history under the treatment sequence $\bar{\ba}_{k-1}$ and $Y^*(\bar{\ba})$ denote the potential outcome under the treatment assignment $\bar{\ba}$.
The set of all potential outcomes is then denoted as $\bW^* = \{ \bH_2^*(a_{1}), \bH_3^*(\bar{\ba}_{2}), \dots, \bH_K^*(\bar{\ba}_{K-1}), Y^*(\bar{\ba}) : \bar{\ba} \in \{0,1\}^K \} $.
The potential outcome of following a regime, $\bd$, is defined as
$$
Y^*(\bd) = \sum_{\bar{\ba} \in \{0,1\}^K} Y^*(\bar{\ba}) \mathbb{I}\{d_1(\bh_1) = a_1\} \prod_{k=2}^{K} \mathbb{I}\left[d_k\{\bH_k^*(a_{k-1})\} = a_k \right]
$$
where $\mathbb{I}$ is the indicator function. Define the value of any regime by $V(\bd) = \mathbb{E}\{ Y^*(\bd) \}$. An optimal regime, $\bd^{opt}$, is then defined as a regime that satisfies
$ V(\bd^{opt}) \geq V(\bd) $ for all $\bd$.

To be able to express the optimal regime in terms of the observed data we will need three standard causal assumptions for DTRs \citep{robins2004optimal}: (C1) the stable unit treatment value assumption (SUTVA), $Y = Y^*(\bar{\bA})$ and $\bH_k = \bH_k^*(\bar{\bA}_{k-1})$ for $k=2,\dots,K$; (C2) sequential ignorability, $ \bW^* \perp A_k | \bH_k $ for $k=1,\dots,K$; and (C3) positivity, $P(A_k = a_k | \bH_k = \bh_k) > 0$ with probability one for each $a_k \in \{0,1\}$ for $k=1,\dots,K$. 

Many estimation methods for an optimal regime focus on estimating a contrast function for how the interaction of treatment and patient history effects the outcome of interest. These estimation methods are commonly referred to as A-learning or advantage learning methods \citep{blatt_2004}. The optimal blip-to-zero function, $\gamma_k(\bh_k, a_k)$, is defined as the difference in expected outcome between receiving treatment $a_k$ and some reference treatment, which we will take to be treatment 0,
for a patient that has history $\bH_k = \bh_k$ if we assume they are treated optimally after stage $k$. Therefore we have that
\begin{align*}
	&\gamma_K(\bh_K, a_K) = \mathbb{E} \{Y^*(\bar{\ba}_{K-1}, a_K)- Y^*(\bar{\ba}_{K-1}, 0) | \bH_K = \bh_K \}, \\
	&\gamma_k(\bh_k, a_k) = \mathbb{E} \left\{  Y^*\left( \bar{\ba}_{k-1}, a_k, \underline{\bd}_{k+1}^{opt} \right)  - Y^* \left(\bar{\ba}_{k-1}, 0, \underline{\bd}_{k+1}^{opt} \right) \vert  \bH_k = \bh_k \right\} ~\mathrm{for}~ k = 2,\dots,K-1, \\
	&\gamma_1(\bh_1, a_1) = \mathbb{E} \left\{  Y^*\left( a_1, \underline{\bd}_{2}^{opt} \right) - Y^* \left( 0, \underline{\bd}_{2}^{opt} \right) \vert  \bH_1 = \bh_1 \right\}.
\end{align*}
Note that the treatments assigned by $\underline{\bd}_{k+1}^{opt}$ may differ according to its arguments. E.g., consider $\gamma_k(\bh_k, a_k)$: for the term $Y^*\left( \bar{\ba}_{k-1}, a_k, \underline{\bd}_{k+1}^{opt} \right)$ we will have that $d_{k+1}^{opt}$ assigns treatment $d_{k+1}^{opt}\{\bH_{k+1}^*(\bar{\ba}_{k-1}, a_k)\} $ while in $Y^*\left(\bar{\ba}_{k-1}, 0, \underline{\bd}_{k+1}^{opt} \right)$ we have that $d_{k+1}^{opt}$ assigns treatment $d_{k+1}^{opt}\{\bH_{k+1}^*(\bar{\ba}_{k-1}, 0)\} $. Then an optimal treatment can be seen as the treatment that maximizes the blip function at that stage, so we have that $ d_k^{opt}(\bh_k) = \arg\max_{a_k} \gamma_k(\bh_k, a_k) $.

A-learning methods alternatively can focus on the regret function. The regret function at stage $k$, $\mu_k(\bh_k, a_k)$, is defined as the decrease in expected outcome from assigning treatment $a_k$ instead of the optimal treatment if we assume that they were treated optimally in all following stages. Therefore for the $K$ stages, the regret functions are given by
\begin{align*}
	& \mu_K(\bh_K, a_K) = E\{Y^*(\bar{\ba}_{K-1},  d_K^{opt})  - Y^*(\bar{\ba}_{K-1}, a_K) | \bH_K = \bh_K \} , \\
	& \mu_k(\bh_k, a_k) = E\{Y^*(\bar{\ba}_{k-1},  d_{k}^{opt}, \underline{\bd}_{k+1}^{opt})  - Y^*(\bar{\ba}_{k-1},  a_{k}, \underline{\bd}_{k+1}^{opt})| \bH_k = \bh_k \}  ~\mathrm{for}~ k= 2,\dots,K-1, \\
	& \mu_1(\bh_1, a_1) = E\{Y^*(d_1^{opt},  \underline{\bd}_2^{opt})  - Y^*(a_1, \underline{\bd}_2^{opt}) | \bH_1 = \bh_1 \}.
\end{align*}
The regret function and the optimal blip-to-zero function are then related by $\mu_k(\bh_k, a_k) = \allowbreak \gamma_k\{ \bh_k, d_k^{opt}(\bh_k) \} -  \gamma_k(\bh_k, a_k)$.
The value of an optimal regime can then be expressed as
\begin{align*}
	V(\bd^{opt}) 
	= \mathbb{E} \left( Y^*(\bar{\bA}) + \sum_{k=1}^{K}
	\left[  \gamma_k \{ \bH_k, d_k^{opt}(\bH_{k}) \} - \gamma_k(\bH_k, A_k) \right] \right).
\end{align*}

Our  sample size procedure is based on using dWOLS to estimate the parameters of the blip functions which then leads to a substitution estimator for the value of the optimal regime. There are two conditions required for
our sample size estimator. These conditions match those used in \cite{Rose_SMART} which were based on those used in \cite{laber2016using}. We will let an estimator of
$\bd^{\mathrm{opt}}$ from an observational study of size $n$ be denoted by $\widehat{\bd}_{n}$. Let $B_0 \in \mathbb{R}$ be a fixed, known mean value such that we want to test whether assigning treatment by following $\bd^{\mathrm{opt}}$ would lead to a mean outcome greater than $B_0$. The choice of $B_0$ will depend on the research question of interest and could represent the mean outcome under a specific static or dynamic regime, e.g. a fixed treatment sequence or standard of care. Let $\eta, \epsilon > 0$,
$\phi, \alpha, \zeta \in
(0,1)$ be constants.  Our goal is to choose $n$ such that:
\begin{itemize}
	\item[(PWR)] there exists an $\alpha$-level test of the null hypothesis, $H_0:
	V(\bd^{\mathrm{opt}})  \le B_0$, based on the estimator,
	$\widehat{\bd}_{n}$, that has a power of at least
	$(1-\phi)\times 100 + o(1)$
	provided
	$V(\bd^{\mathrm{opt}}) \ge B_0 + \eta$;
	\item[(OPT)] $P\{V(\widehat{\bd}_n) \geq V(\bd^{\mathrm{opt}}) - \epsilon\} \geq 1 - \zeta + o(1)$.
\end{itemize}
Each of these conditions provide guarantees on different aspects of our sample size procedure. The first condition, (PWR), ensures that if tailoring treatments based on patient history provides a \emph{clinically} significant improvement in outcomes, then a study sized with our approach will be sufficiently powered to detect that there is a \emph{statistically} significant difference between the value of the optimal regime and $B_0$. The second condition, (OPT), guarantees the true value of a regime estimated from a study based on our power calculations will be within a specified tolerance of the value of the true optimal regime. The quantity $V(\widehat{\bd}_n)$ represents the marginal mean outcome if the estimated optimal regime $\widehat{\bd}_n$ was used to assign treatments to the population of interest and can be expressed as $V(\widehat{\bd}_n) = \mathbb{E}\{Y^*(\widehat{\bd}_n) | \mathcal{D}_n\} $.
This will ensure that the performance of our estimated optimal regime will be close to the true optimal regime.

\section{Methodology of Proposed Sample Size Calculations}
\label{s: methodology}

\subsection{Dynamic Weighted Ordinary Least Squares}

Dynamic weighted ordinary least squares estimates parameters in the blip functions using a sequence of weighted ordinary least squares regressions \citep{Wallace_2015}. We posit a model for the blip-to-zero functions given by $\gamma_k(\bh_k, a_k; \psi_k)$ for $k=1\dots,K$. We will assume that each blip-to-zero function is correctly specified such that $\gamma_k(\bh_k, a_k; \psi_k^*) = \gamma_k(\bh_k, a_k)$.

The treatment-free outcome is given by
$$
G_k(\underline{\psi}_k) = Y - \gamma_k(\bh_k, a_k; \psi_k)  + \sum_{j=k+1}^{K} \mu_j(\bh_j, a_j; \psi_j).
$$
This represents the patients' actual outcome adjusted for the expected difference in a patients' outcome if they were to receive treatment 0 at stage $k$ and then were treated optimally for the remaining stages if we assume that $\underline{\psi}_k$ is the true value of the parameter in our blip models. This is referred to as the treatment-free outcome since it
does not depend on the treatment received at stage $k$, though it is only ``treatment-free'' if the reference treatment is no treatment. When we are considering only active treatments, the treatment-free outcome denotes the expected outcome under the reference treatment at stage $k$.
Under assumptions (C1)-(C3), we have that $\mathbb{E}\{G_k(\underline{\psi}_k^*)|\bH_k = \bh_k \} = \mathbb{E}\{Y^*(\bar{\ba}_{k-1}, 0, \underline{\bd}^{\mathrm{opt}}_{k+1}) | \bH_k = \bh_k \}$.
We then specify a model for $\mathbb{E} \{G_k(\underline{\psi}_k^*) | \bH_k = \bh_k\}$ which will be given by $g_k(\bh_k; \beta_k)$.

Define the pseudo-outcome for stage $k$ as
$
\tilde{Y}_k = Y + \sum_{j=k+1}^{K} \mu_j(\bh_j, a_j; \hat{\psi}_j).
$
The pseudo-outcome for stage $k$ represents the estimated counterfactual outcome if treatments were assigned via our estimated optimal rules after stage $k$. We model the pseudo-outcome as the sum of the treatment-free model and the blip-to-zero model
$$
\mathbb{E}\left(\tilde{Y}_k | \bH_k = \bh_k, A_k = a_k; \beta_k, \psi_k \right) = g_k(\bh_k; \beta_k) + \gamma_k(\bh_k, a_k; \psi_k).
$$
We could then estimate $\beta_k$ and $\psi_k$ as a standard regression problem which would lead to an estimated optimal decision rule at stage $k$,
$
\hat{d}_k^{opt}(\bh_k) = \arg\max_{a_k \in \{0,1\}} \gamma_k (\bh_k, a_k; \hat{\psi}_k).
$
Note that the estimated regime only depends on $\psi_k$ so $\beta_k$ is a nuisance parameter, but to have a consistent estimator of $\psi_k$ would require correctly specifying our models for both $g_k(\bh_k; \beta_k)$ and $\gamma_k(\bh_k, a_k; \psi_k)$. Therefore we posit a model $\pi_k(\bh_k; \xi_k)$ for the propensity score $\pi_k(\bh_k) = \mathbb{E}(A_k | \bH_k = \bh_k)$. \cite{Wallace_2015} show that if we perform a weighted ordinary least squares with a weight function that satisfies
$
\pi_k(\bh_k; \xi_k)w(1, \bh_k; \xi_k) = \{1 - \pi(\bh_k; \xi_k) \} w(0, \bh_k; \xi_k)
$
then the resulting estimate of $\psi_k$ will be consistent as long as the blip model is correctly specified and at least one of the treatment-free model or the propensity score model is correctly specified. Weight functions that satisfy this equality include $w(a_k, \bh_k; \xi_k) = |a_k - \pi(\bh_k; \xi_k)|$ and the inverse probability of treatment weights given by
$
w(a_k, \bh_k; \xi_k) = a_k \{\pi(\bh_k; \xi_k)\}^{-1} + (1 - a_k)\{1 - \pi(\bh_k; \xi_k)\}^{-1}.
$

\subsection{Inference for the Value}

We assume linear models for both the treatment-free and blip models.
Therefore the model for the pseudo-outcome will be given by
\begin{align*}
	\mathbb{E} \left(\tilde{Y}_k | \bH_k = \bh_k, A_k = a_k; \beta_k, \psi_k  \right) 
	& = \bh_{k,\beta}^T \beta_k + a_k \bh_{k,\psi}^T \psi_k
\end{align*}
where $\bh_{k,\beta}$ and $\bh_{k,\psi}$ are components of $\bh_{k}$, each including a leading one. The estimated optimal treatment at stage $k$ is given by $\mathbb{I}(\bh_{k,\psi}^T \hat{\psi}_k > 0)$ and the pseudo outcome for stage $k$ is given by
$
\tilde{Y}_k 
= Y + \sum_{j=k+1}^{K}  \bh_{j,\psi}^T \hat{\psi}_j \left\{ \mathbb{I}(\bh_{j,\psi}^T \hat{\psi}_j > 0) - a_j \right\}.
$
Note the pseudo-outcome at stage $k$ is a non-smooth function of the generative model because of the indicator function. Therefore the estimator for $\psi_k$ is non-regular when $k < K$ and standard approaches for inference no longer hold because $\sqrt{n}\left(\hat{\psi}_k - \psi_k \right)$ is not uniformly normal \citep{robins2004optimal}.

Similarly, an estimator for the value is given by
\begin{align*}
	\widehat{V}\left(\widehat{\bd}^{opt}\right) 
	& = \mathbb{P}_n \left[ Y + \sum_{k=1}^{K}  \bH_{k,\psi}^T \hat{\psi}_k \left\{ \mathbb{I}(\bH_{k,\psi}^T \hat{\psi}_k > 0) - A_k \right\} \right]
\end{align*}
where $\mathbb{P}_n$ denotes the empirical expectation. $V\left(\bd^{opt}\right)$ is a non-smooth function of the generative model as well so we again have that standard approaches for inference do not hold. Therefore, to derive a test that satisfies our (PWR) sample size condition, we invert a projection confidence interval for $V\left(\bd^{opt}\right)$ \citep{laber_2014}. This interval is valid as long as the blip model is correctly specified in addition to either the propensity score model or the treatment-free model being correctly specified.

Define $Y(\psi) =  Y + \sum_{k = 1}^{K} \mu_k(\bH_k, A_k; \psi_k)$ and $\widehat{V}_n(\psi) = \mathbb{P}_n \{ Y(\psi) \}$. Thus $\widehat{V}_n(\hat{\psi}) = \widehat{V}(\widehat{\bd}^{opt})$. We also have that $V(\bd^{opt}) = \mathbb{E} \{ Y(\psi^*) \}$. Define $\varsigma^2(\psi) = \mathbb{E}\left\{Y(\psi) - \mathbb{E}Y(\psi) \right\}^2 $ and $\hat{\varsigma}_n^2(\psi) = \mathbb{P}_n\left\{Y(\psi) - \mathbb{P}_nY(\psi) \right\}^2 $. Then for a fixed value of $\psi$, if $\mathbb{E}\{ Y^2(\psi) \} < \infty$
$$
\sqrt{n}\left\{ \widehat{V}_n(\psi) - V(\psi) \right\}
\rightarrow \mathrm{Normal}\{0, \varsigma^2(\psi) \}.
$$
Let $\Psi_{n, 1 - \vartheta}$ denote a $(1-\vartheta) \times 100\%$ confidence region for $\psi^*$. If we choose $\vartheta_1$ and $\vartheta_2$ such that $\vartheta_1 + \vartheta_2 = \alpha$, then an $\alpha$-level test for (PWR) rejects when
$$
\inf_{\psi \in \Psi_{n, 1 - \vartheta_1}} \left\{
\widehat{V}_n(\psi) - \frac{z_{1-\vartheta_2}\hat{\varsigma}_n(\psi)}{\sqrt{n}}
\right\} \geq B_0.
$$
See Section A of the online supplementary material for proof that this is an $\alpha$-level test. The power for this test is given by
\begin{align*}
	& P\left[
	\inf_{\psi \in \Psi_{n, 1 - \vartheta_1}} \left\{
	\widehat{V}_n(\psi) - \frac{z_{1-\vartheta_2}\hat{\varsigma}_n(\psi)}{\sqrt{n}}
	\right\} \geq B_0	
	\right] \\
	& \quad \geq P\left\{
	\inf_{\psi \in \Psi_{n, 1 - \vartheta_1}}
	\left[
	\frac{\sqrt{n}\{ \widehat{V}_n(\psi) - V(\psi) \}}{\hat{\varsigma}_n(\psi) }
	+ \frac{\min \left[ \sqrt{n}\{V(\psi) - B_0\} , \sqrt{n}\eta \right]}{\hat{\varsigma}_n(\psi)}
	\right] \geq z_{1-\vartheta_2}
	\right\}.
\end{align*}
We replace $V(\psi) - B_0$ with $\min\{V(\psi) - B_0, \eta\}$ as in \cite{Rose_SMART} so that the sample size is based on the minimal effect size of interest instead of the estimated effect size. This will result in our proposed sample size procedure as having power $(1 - \phi) \times 100 \%$ when the effect size is $\eta$ and the power will increase as the true effect size increases to greater than $\eta$.

\subsection{Confidence Region for $\psi$ }

The proposed hypothesis test requires constructing a confidence region for $\psi$. When $K=1$ constructing a confidence region for $\psi$ can be done using standard theory for m-estimators \citep{van_der_vaart_book}. Let $\bH_{k,\beta}$ be the components of the history in the treatment-free model at stage $k$ and let $\bH_{k,\psi}$ be the components of the history in the blip-to-zero model at stage $k$. For $K=1$, the joint estimating equations are given by
\begin{align*}
	\sum_{i = 1}^{n}
	\begin{pmatrix}
		\bH_{1,\beta} \\
		A_1 \bH_{1,\psi}
	\end{pmatrix} w_1(\bH_1, A_1; \xi_1)(Y - \bH_{1,\beta}^T\beta_1 - A_1\bH_{1,\psi}^T\psi_1) = 0 \\
	\sum_{i = 1}^{n}
	\begin{pmatrix}
		1 \\ \bH_1
	\end{pmatrix}
	\left\{
	A_1 - \frac{\exp(\xi_{11} + \bH_1^T\xi_{12}) }{1 + \exp(\xi_{11} + \bH_1^T\xi_{12})}
	\right\} = 0.
\end{align*}

The standard sandwich variance estimator that does not adjust for the propensity score estimation performs well in practice \citep{wallace_2017}. Denoting the variance estimator by  $\Sigma_{\hat{\psi}_1}$, $\mathfrak{Z}_{\epsilon} = \{\psi_1: n(\psi_1 - \hat{\psi}_1) \Sigma_{\hat{\psi}_1}^{-1}(\psi_1 - \hat{\psi}_1) \leq \chi^2_{1-\epsilon, p_1} \}$  is a Wald-type asymptotic $(1-\epsilon)\times 100\%$ confidence region for $\psi_1$.

When $K > 1$, as previously mentioned, the estimator for $\psi_k$ when $k < K$ is non-regular due to the non-smoothness of the pseudo-outcome. 
One potential solution to construct a valid confidence set for $\psi$ is through the use of a projection interval.
Define
$
\tilde{Y}_k\left(\underline{\psi}_{k+1}\right) 
= Y + \sum_{j=k+1}^{K}  \bh_{j,\psi}^T \psi_j \left\{ \mathbb{I}(\bh_{j,\psi}^T \psi_j > 0) - a_j \right\}
$
so we have that $\tilde{Y}_k(\underline{\psi}_{k+1})$
is equivalent to the pseudo-outcome at stage $k$ if $\underline{\psi}_{k+1} = \underline{\hat{\psi}}_{k+1}$.
For $k = 1, \dots , K-1$ define
$$
\psi_k^*(\underline{\psi}_{k+1}) = \arg\min_{\psi_k} \mathbb{E} \left[ w_k(\bH_k, A_k) \left\{\tilde{Y}_k\left(\underline{\psi}_{k+1}\right) - \bH_{k,\beta}^T\beta_k - A_k\bH_{k,\psi}^T\psi_k \right\}^2
\right]
$$
so that $\psi_k^*(\underline{\psi}_{k+1})$ denotes the population-level parameter for the blip-to-zero model if we knew that $\underline{\psi}^*_{k+1} = \underline{\psi}_{k+1}$. Therefore we also have that $\psi^*_k = \psi_k^*(\underline{\psi}_{k+1}^*)$. Define an estimator for $\psi_k^*(\underline{\psi}_{k+1})$ to be given by
$$
\hat{\psi}_k(\underline{\psi}_{k+1}) = \arg\min_{\psi_k} \mathbb{P}_n \left[ w_k(\bH_k, A_k; \hat{\xi}) \left\{\tilde{Y}_k\left(\underline{\psi}_{k+1}\right) - \bH_{k,\beta}^T\beta_k - A_k\bH_{k,\psi}^T\psi_k \right\}^2
\right].
$$
This estimator is the weighted least squares estimator used in dWOLS with the pseudo-outcome replaced with $\tilde{Y}(\underline{\psi}_{k+1})$.
Let $\Sigma_{\hat{\psi}_k}(\underline{\psi}_{k+1}) $ denote the variance of $\hat{\psi}_k(\underline{\psi}_{k+1})$. Then
$$
\mathfrak{Z}_{k,n,\epsilon}(\underline{{\psi}}_{k+1})  = \left\{\psi_k: n(\psi_k - \hat{\psi}_k) \Sigma_{\hat{\psi}_k}^{-1}(\underline{\psi}_{k+1} )(\psi_k - \hat{\psi}_k) \leq \chi^2_{1-\epsilon, p_k} \right\}
$$
gives a $(1-\epsilon)\times 100\%$ Wald-type asymptotic confidence region for $\psi_k^*(\underline{\psi}_{k+1})$.
Then given $\epsilon_1, \dots, \epsilon_K \in (0,1)$ such that $\vartheta = \sum_{k=1}^{K} \epsilon_K \leq 1$
$$
\Psi_{n,\vartheta} = \left\{
\psi: \psi_K \in \mathfrak{Z}_{K,n,\epsilon_K} ~\mathrm{and}~ \psi_k \in \mathfrak{Z}_{k,n,\epsilon_k}(\underline{\psi}_{k+1}) ~\mathrm{for}~ k=1,\dots,K-1
\right\}.
$$
represents a $(1 - \vartheta)\times 100\%$  confidence region for $\psi^*$.
This approach to creating a confidence region for $\psi$ will be increasingly conservative as $K$ increases. Also, recall that our hypothesis test involves finding
$ \inf_{\psi \in \Psi_{n, 1 - \vartheta_1}} \left\{
\widehat{V}_n(\psi) - \frac{z_{1-\vartheta_2}\hat{\varsigma}_n(\psi)}{\sqrt{n}}
\right\}
$
which will be computationally difficult for large values of $K$.

Alternatively, we can construct a valid confidence set using the $m$-out-of-$n$ bootstrap. The  $m$-out-of-$n$ bootstrap is a tool for producing valid confidence intervals for non-smooth functionals \citep{Swanepoel_1986, Dumbgen_1993, Shao_1994, Bickel_1997}. The $m$-out-of-$n$ bootstrap uses a resampling size $m$ that is smaller than the sample size $n$. \cite{Bibhas_2013} used the $m$-out-of-$n$ bootstrap
to create valid confidence intervals for the parameters indexing a dynamic treatment regime when estimating an optimal regime using Q-learning as well as an adaptive method to select $m$. \cite{Simoneau_2018} examined using this procedure to create valid confidence intervals when using dWOLS to estimate the optimal regime. Both of these papers focused on two-stage dynamic treatment regimes so that the estimator for only the first stage has a non-regular limiting distribution. We propose a method for generalizing this procedure to a $K$-stage dynamic treatment regime in which the estimators in all $k=1\dots,K-1$ stages suffer from non-regularity.

We first discuss how the $m$-out-of-$n$ bootstrap is used to construct confidence intervals for stage one parameters for a two-stage dynamic treatment regime. Define $p \triangleq P(\bH_{2,\psi}^T\psi_2 = 0)$, so we have that $p$ is a measure of the degree of non-regularity in the data. When $p = 0$, the distribution of $\sqrt{n}(\hat{\psi}_1 - \psi_1) $ is asymptotically normal and the standard bootstrap will produce valid confidence intervals. 
\cite{Bibhas_2013} proposed using a resample size of
$
\hat{m} \triangleq n^{\frac{1 + \kappa(1-\hat{p})}{1 + \kappa}}
$
where $\kappa > 0$ is a tuning parameter and $\hat{p}$ is an estimate of $p$. When $\hat{p} = 0$, we have that $m=n$ and as $\hat{p}$ increases our resample size will decrease while $\kappa$ determines the smallest acceptable resample size with $m$ taking values within the interval $[n^{\frac{1}{1 + \kappa}}, n]$. They proposed using a plug-in estimator for $p$ given by
$ \hat{p} = \mathbb{P}_n \mathbb{I}\{n(\bH_{2,\psi}^T\hat{\psi}_{2})^2 \leq \tau_n(\bH_{2,\psi}) \} $ where $\tau_n(\bh_{2,\psi})$ is given by $\bh_{2,\psi}^T\hat{\Sigma}_{\hat{\psi}_2}\bh_{2,\psi}\chi^2_{1,1-\nu}$ such that $\hat{\Sigma}_{\hat{\psi}_2}$ is the plug-in estimator of $n \cov (\hat{\psi}_2, \hat{\psi}_2)$. Let $\hat{\psi}_{1,\hat{m}}^{(b)}$ denote the bootstrap estimate for $\psi_1$ from using a resample size of $\hat{m}$. To construct a $(1 - \vartheta) \times 100 \%$ confidence interval for $\psi_1$, calculate the $\vartheta/2 \times 100$ and $(1 - \vartheta/2) \times 100$ percentiles of $\sqrt{\hat{m}}(\hat{\psi}_{1,\hat{m}}^{(b)} - \hat{\psi}_1)$ which we will denote by $\hat{l}$ and $\hat{u}$ respectively. Then a $(1 - \vartheta) \times 100 \%$ confidence interval for $\psi_1$ is given by $(\hat{\psi}_1 - \hat{u}/\sqrt{\hat{m}}, \hat{\psi}_1 - \hat{l}/\sqrt{\hat{m}})$.

To generalize to a $K$ stage dynamic treatment regime, we will start by defining $p_k \triangleq P(\bH_{k+1,\psi}^T\psi_{k+1} = 0)$. Therefore $p_k$ indicates the degree of non-regularity in the estimation of $\psi_k$ at stage $k$. The plug-in estimator for $p_{K-1}$ can be calculated using $\hat{\Sigma}_{\hat{\psi}_K}$. For $p_k$ for $k = 1, \dots, K-2$ the non-regularity will cause the usual plug-in estimator to no longer be valid. Instead we will use the $m$-out-of-$n$ bootstrap to construct a valid confidence interval for $\bh_{k,\psi}^T\psi_k$. The estimator $\hat{p}_k$ for $p_k$ is then given by the proportion of individuals in the sample for which this confidence interval for $\bh_{k,\psi}^T\psi_k$ contains zero. We then move backwards through the stages getting an estimate of $\hat{p}_k$ at each stage using the $m$-out-of-$n$ bootstrap with a resample size of $\hat{m}_k = n^{\frac{1 + \kappa(1-\hat{p}_k)}{1 + \kappa}}$ at each stage. Let $\hat{p} = \max_{k} \hat{p}_k$. We calculate $\hat{m}$ using the same formula as before and use this as our resample size. Then calculate the $\epsilon_k/2 \times 100$ and $(1 - \epsilon_k/2) \times 100$ percentiles of $\sqrt{\hat{m}}(\hat{\psi}_{k,\hat{m}}^{(b)} - \hat{\psi}_k)$ which we will denote by $\hat{l}_k$ and $\hat{u}_k$ for each value of $k$.
Then given $\epsilon_1, \dots, \epsilon_K \in (0,1)$ such that $\vartheta = \sum_{k=1}^{K} \epsilon_K \leq 1$, a $(1 - \vartheta)\times 100\%$  confidence region for $\psi^*$ is given by
$$
\Psi_{n,\vartheta} = \{\psi:  \psi_k \in (\hat{\psi}_k - \hat{u}_k/\sqrt{\hat{m}}, \hat{\psi}_k - \hat{l}_k/\sqrt{\hat{m}}) ~\mathrm{for}~ k = 1,\dots, K-1 ~\mathrm{and}~ \psi_K \in \mathfrak{Z}_{K,n,\epsilon_K} \}. $$
Section B of the online supplementary material contains simulations demonstrating the coverage of confidence intervals generated using this procedure when applied to $K = 3$ stage dynamic treatment regimes and a data-driven approach to selecting $\kappa$ using the double-bootstrap.

\subsection{Bootstrap Power Calculations}

Our sample size procedure is based on estimating the power for a given sample size using a bootstrap of pilot data, i.e. resampling from the pilot data samples of size $n$ and assessing power with that sample size over a grid of candidate sample sizes $n \in \mathbb{N}$. We assume pilot data $\mathcal{D}_{n_0} = \{(\bX_{1,i}, A_{1,i}, \dots, \bX_{K,i}, \allowbreak  A_{K,i}, Y_i) \}_{i=1}^{n_0}$ that is comprised of $n_0$ $i.i.d$ replicates from the same population of interest as the full study. We estimate the power for a given sample size using the bootstrap and search for the smallest sample size $n$ that exceeds the threshold given in condition (PWR).

Let $P_B$ denote probabilities computed with respect to the bootstrap distribution conditional on the pilot data. The bootstrap estimator of the minimum sample size required to satisfy condition (PWR) is given by the smallest $n$ that satisfies
\begin{multline*}
	P_B\left\lbrace\rule{0cm}{1.15cm}\right.
	\inf_{\psi \in \Psi_{n_0, n, 1-\vartheta_1}^{(b)}}
	\left(\rule{0cm}{1.05cm}\right.
	\frac{
		\sqrt{n}\left\lbrace
		\widehat{V}_{n_0,n}^{(b)}(\psi) - \widehat{V}_{n_0}(\psi)
		\right\rbrace
	}{
		\widehat{\varsigma}_{n_0,n}^{(b)}(\psi)
	}
	+ \\
	\frac{
		\min\left[
		\sqrt{n}\left\lbrace
		\widehat{V}_{n_0}(\psi) - B_0
		\right\rbrace,
		\,\sqrt{n}\eta\right]
	}{
		\widehat{\varsigma}_{n_0,n}^{(b)}(\psi)
	}
	\left. \rule{0cm}{0.95cm}\right)
	\ge z_{1-\vartheta_2}
	\left. \rule{0cm}{1.05cm}\right\rbrace \ge 1-\gamma
\end{multline*}
such that $\vartheta_1 + \vartheta_2 = \alpha$. \cite{Rose_SMART} proved that a bootstrap oversampling estimator of this form is consistent as $n_0$ and $n$ diverge under mild assumptions. Here the form of $\widehat{V}_{n_0}(\psi)$ is different, which leads to having to make slightly different assumptions. If we assume:
\begin{itemize}
	\item[(A1)] $\inf_{\psi} \mathbb{E} \left[
	Y(\psi) - \mathbb{E}\{Y(\psi)\} \right]^2 > 0$ and $\sup_{\psi} \mathbb{E} \left[
	Y(\psi) - \mathbb{E}\{Y(\psi)\} \right]^2 < \infty$;
	\item[(A2)] the classes $\mathcal{F}_1 \{ Y(\psi) : \psi \in \Theta\}$ and $\mathcal{F}_2 = \{Y^2(\psi): \psi \in \Theta \}$ are Donsker;
	\item[(A3)] $\mathbb{E} \{ Y(\psi) \}$  is uniformly continuous in a neighborhood of $\psi^*$;
\end{itemize}
we have that consistency holds and the proof then follows that of \cite{Rose_SMART}.

Now we focus on determining sample sizes for the (OPT) condition. Recall, this condition states that $P\{V(\widehat{\bd}_n) \geq V(\bd^{\mathrm{opt}}) - \epsilon\} \geq 1 - \zeta + o(1)$. Note that for any sequence $\tilde{\psi}_n \in \Psi_{n,1 -\vartheta}$ such that $\widehat{V}_n(\psi^*) \leq \widehat{V}_n(\tilde{\psi}_n) + o_P(1/\sqrt{n})$ we have that
\begin{align*}
	P\left[
	V(\tilde{\psi}) \geq V(\bd^{opt}) + \inf_{\psi \in \Psi_{n, 1-\vartheta_1}} \{\widehat{V}(\psi) - V(\psi) \} - \sup_{\psi \in \Psi_{n, 1-\vartheta_1}} \{\widehat{V}(\psi) - V(\psi) \} 	
	\right] \geq 1 - \vartheta_1 + o(1).
\end{align*}
Then if $\mathfrak{Q}_{n,1-\vartheta_2, 1-\vartheta_1}$ is the $(1-\vartheta_2)$th quantile of
$$
\sup_{\psi \in \Psi_{n, 1-\vartheta_1}} \{\widehat{V}(\psi) - V(\psi) \} ~~ - \inf_{\psi \in \Psi_{n, 1-\vartheta_1}} \{\widehat{V}(\psi) - V(\psi) \} 	
$$
we have that (OPT) holds asymptotically if $\vartheta_1 + \vartheta_2 \leq \zeta$ and $\mathfrak{Q}_{n,1-\vartheta_1, 1-\vartheta_2} \leq \epsilon$. We again use bootstrap oversampling to estimate the smallest $n$ such that this holds. Let $\mathfrak{Q}_{n_0, n, 1-\vartheta_2, 1-\vartheta_1}^{(b)}$ be the bootstrap estimate of $\mathfrak{Q}_{n,1-\vartheta_2, 1-\vartheta_1}$  from a pilot study of size $n_0$ with a resample size of $n$. Then the estimate of our sample size is given by the smallest $n$ such that $\mathfrak{Q}_{n_0, n, 1-\vartheta_2, 1-\vartheta_1}^{(b)} \leq \epsilon$. To calculate a sample size that satisfies both conditions simultaneously, we recommend calculating a sample size for each condition individually and use the max of the two.

\section{Simulation Study}
\label{s: simulation}

We examined the finite sample performance of our proposed sample size procedure with a simulation study. We considered sizing a three-stage study with two treatment options at each stage. To evaluate the performance of our method we conducted simulations for each of the two conditions (PWR) and (OPT) individually. Section C of the online supplementary material contains additional simulations for sizing a two-stage study.
The data generating model for our simulations was:
\begin{equation*}
	\begin{array}{ll}
		X_1 \sim \mathrm{N}\{ 0, 1\},  
		& P(A_k = 1| \bH_k = \bh_k) = \left\{1 + e^{-(\varpi_{k,0} + \varpi_{k,1} x_k )}\right\}^{-1}~\mathrm{for}~ k = 1,2,3,\\
		\tau_1 \sim \mathrm{N}(0, 1), &  X_2 = \mu_{20} +  \mu_{21} X_1 + \tau_1, \\ 
		\tau_2 \sim \mathrm{N}(0, 1), &  X_3 = \mu_{30} + \mu_{31} X_1 + \mu_{32} X_2 + \tau_2, \\ 
		\bH_{3,1}^\T = (1, X_1, X_2, X_3), & \bH_{3,0}^\T = (1, X_1, A_1, A_1X_{1}, X_2, A_2, A_2 X_1, A_2 X_2, X_3, X_1^2)  \\	
		\upsilon \sim \mathrm{N}(0, 1),
		& Y = \bH_{3,0}^\T \lambda_{3,0} + A_3\bH_{3,1}^\T \lambda_{3,1} + \upsilon. \\
	\end{array}
\end{equation*}
The parameters of the data generating model were given by: 
\begin{align*}
	\begin{array}{ll}
		\varpi_1 = (0.25, 1), &
		\varpi_2 = (0.25, 1, -1, -1), \\
		\varpi_3 = (0.25, 0.5, 0.5, -0.5, 1, -0.5), &  \\
		\mu_2 = (0, 0.5), & \mu_3 = (0, -0.5, 0.5), \\
		\lambda_{3,0} = (1, 1, 0.5, -0.75, 0.5, -0.5, -0.5, 0.5, 0.5, 0.25),  & \lambda_{3,1} = (0.25, 0.5, 0.5, -0.5). 
	\end{array}
\end{align*}

We posited models such that the blip model at each stage was correctly specified, but both treatment-free models were misspecified by leaving out $X_{1}^2$. We modeled the propensity score with a correctly specified logistic regression model so that dWOLS produced consistent estimates of the blip parameters.

We assumed $\alpha = 0.05$, $\phi = 0.1$, and $\eta  = 1.4$. Therefore, the first condition (PWR) held if we had a 0.05 level test of $H_0: V(\bd^{\mathrm{opt}})  \le B_0$ that had power 90\% provided $V(\bd^{\mathrm{opt}}) \ge B_0 + 1.4$. We evaluated the performance of the sample size procedure when the effect size of tailoring was equal to $\eta = 1.4$ and examined how it changed as the effect size increased. An effect size of $\eta = 1.4$ corresponded to a standardized effect size of 0.72, which is relatively moderate \citep{cohen_1992}. We let $V(\bd^{\mathrm{opt}}) = B_0 + \eta + \Delta\eta$ and varied $\Delta \in \{0, 0.5, 1, 1.5\}$. The data generating model was fixed across all settings which caused $V(\bd^{\mathrm{opt}})$ to be fixed, so we let $B_0$ vary with $\Delta$ such that $B_0 = V(\bd^{\mathrm{opt}}) - \Delta\eta - \eta$. We let $\zeta = 0.1$ and varied $\epsilon \in \{0.3, 0.5, 0.7\}$. Therefore the second condition (OPT) held if $P\{V(\widehat{\bd}_n) \geq V(\bd^{\mathrm{opt}}) - \epsilon \} \geq 0.9$.

Each repetition of the simulation study consisted of the following steps. First, we generated a pilot study of size $n_0 \in \{200, 400\}$. Second, we estimated the power on a grid of potential sample sizes using 500 bootstrap repetitions for each sample size considered. To construct a confident set for $\psi$, we used the $m$-out-of-$n$ bootstrap with $\kappa = 0.2$. Third, we used least squares to regress the estimated power on the potential sample sizes and used the fitted model to estimate the smallest sample size, $\hat{n}(\mathcal{D}_{n_0})$, that achieved the desired power of 90\%. We fit this model using only the tested sample sizes that resulted in an estimated power in a small neighborhood of the targeted power as the power curve will only be approximately linear within a small region.
It is possible that the estimate of the value from the pilot study is less than the comparison mean such that $\widehat{V}_{n_0}(\widehat{\psi}) \leq B_0 $. When this occurs, we won't be able to find a sample size using that pilot that is powered for a comparison with $B_0$, and we define $\hat{n}( \mathcal{D}_{n_0} ) = \infty$. Fourth, for each $\hat{n}( \mathcal{D}_{n_0} ) < \infty$, we generated a study of size $\hat{n}(\mathcal{D}_{n_0})$ and performed a hypothesis test for condition (PWR) calculating the empirical power over the 500 simulations using this process. For condition (OPT), we estimated the optimal regime using the study of size $\hat{n}(\mathcal{D}_{n_0})$ and calculated the true value of the estimated regime using the known data generating functions. Lastly, we checked whether this value was within $\epsilon$ of the value of the true optimal regime. Section D of the online supplementary material contains high-level pseudocode for the simulation study.

\begin{table}[h]
	\caption{
		Empirical power (PWR) using the projection-based sample size procedure at a nominal level of 90 using a pilot study of size $n_0 = 200 $ and $n_0 = 400$. $\Delta$ denotes the difference between the true value of the optimal regime and $B_0$ which is given by $\eta(1 + \Delta)$. $P(\hat{n} = \infty)$ represents the proportion of pilot studies for which $\hat{n}( \mathcal{D}_{n_0} ) = \infty$. The remaining columns give the mean, median, quartiles, and standard deviation of the estimated sample sizes across the 500 simulation repetitions.}
	\begin{center}
		\begin{tabular}{| c c c c c c c c c |} \hline
			$\Delta$ & $n_0$ 
			& $\mathbb{E}(\hat{n}) $ & $\mathrm{Q1}(\hat{n}) $ & $\mathrm{Med}(\hat{n}) $ & $\mathrm{Q3}(\hat{n}) $ & $\mathrm{SD}(\hat{n})$ & $P(\hat{n} = \infty ) $ & PWR  \\ \hline	
			0 & 200 & 640.09 & 307.00 & 490.00 & 780.00 & 511.39 & 0.04 & 74.39 \\ 
			0.5 & 200 & 142.23 & 131.00 & 141.00 & 151.00 & 18.21 & 0.00 & 99.80 \\ 
			1 & 200 & 142.73 & 132.00 & 141.00 & 152.00 & 18.32 & 0.00 & 100.00 \\ 
			1.5 & 200 & 142.44 & 132.00 & 141.00 & 151.00 & 17.97 & 0.00 & 100.00 \\ \hline
			0 & 400 & 678.77 & 410.00 & 571.00 & 801.75 & 408.25 & 0.01 & 82.68 \\ 
			0.5 & 400 & 116.67 & 109.00 & 116.00 & 123.00 & 11.38 & 0.00 & 100.00 \\ 
			1 & 400 & 116.39 & 109.00 & 115.50 & 123.00 & 11.11 & 0.00 & 100.00 \\ 
			1.5 & 400 & 116.39 & 109.00 & 115.50 & 123.00 & 11.11 & 0.00 & 100.00 \\ 
			\hline  
		\end{tabular}
	\end{center}
	\label{3_stage_PWR}
\end{table}

\begin{table}[h]
	\caption{
		Empirical concentration (OPT) using the projection-based sample size procedure at a nominal level of 90 using a pilot study of size $n_0 = 200 $ and $n_0 = 400$. We test whether the true value of the estimated regime is within $\epsilon$ of the true value of the true optimal regime. 
		The remaining columns give the mean, median, quartiles, and standard deviation of the estimated sample sizes across the 500 simulation repetitions.}
	\begin{center}
		\begin{tabular}{| c c c c c c c c |} \hline
			$\epsilon$ & $n_0$ 
			& $\mathbb{E}(\hat{n}) $ & $\mathrm{Q1}(\hat{n}) $ & $\mathrm{Med}(\hat{n}) $ & $\mathrm{Q3}(\hat{n}) $ & $\mathrm{SD}(\hat{n})$ & OPT \\ \hline	
			0.30 & 200 & 1844.93 & 1561.50 & 1796.00 & 2060.00 & 365.42 & 100.00 \\ 
			0.50 & 200 & 754.27 & 663.50 & 748.00 & 836.00 & 128.73 & 100.00  \\ 
			0.70 & 200 & 431.14 & 387.00 & 426.00 & 473.00 & 65.13 & 100.00 \\ \hline
			0.30 & 400 & 1319.70 & 1191.50 & 1294.50 & 1451.75 & 193.43 & 100.00 \\ 
			0.50 & 400 & 543.24 & 500.00 & 526.50 & 586.00 & 69.92 & 100.00 \\ 
			0.70 & 400 & 320.63 & 293.25 & 319.00 & 345.75 & 37.88 & 100.00 \\ 
			\hline  
		\end{tabular}
	\end{center}
	\label{3_stage_OPT}
\end{table}

Tables \ref{3_stage_PWR} and \ref{3_stage_OPT} contain the results for sizing for (PWR) and (OPT) respectively. The comparison to $B_0$ was slightly underpowered when $\Delta = 0$ and $n_0 = 200$ with a power of 74.39\%. This is partly due to the size of the pilot study as we can see that as the pilot size increased to $n_0 = 400$, the power increased to 82.68\%. Increasing the size of the pilot also decreased the variance in the estimated sample size. As $\Delta$ increased, the power increased as expected and converged to 100\%. For $\Delta = 0$, the distribution of $\hat{n}(\mathcal{D}_{n_0})$ was right skewed as expected; generally, pilot studies in which the estimated benefit to tailoring is very small will result in a large estimated sample size. This also caused the variance of the estimated sample size to increase. In general, this will occur when using the proposed method when $\eta$ is small and $V(\bd^{\mathrm{opt}})$ is close to $B_0 + \eta$. As $\Delta$ increased, there was a decline in the degree of skewness and variance in the estimated sample size. Table \ref{3_stage_OPT} shows that the nominal concentration of 90\% was achieved for all values of $\epsilon$ and $n_0$. This procedure is conservative, as we can see that the concentration was 100\% for all simulation settings. As $\epsilon$ decreased, the estimated sample size increased as expected. We also see that as we increasd the size of the pilot study, the variance in the sample size estimate decreased. The simulation results from the two-stage study contained in Section C of the online supplementary material were very similar.

\section{Illustration Using Data Gathered from EHRs}
\label{s: KPWA}

Kaiser Permanente Washington is a health system providing both clinical care and health insurance to members. This study used data extracted from electronic health records and health insurance claims from KPWA clients. The Kaiser Permanente Washington Institutional Review Board approved use of these de-identified data under waiver of consent under approval number 1561019-4. The data used for this study consist of records of 82,691 patients who began antidepressant treatment for depression from 2008 through 2018 and includes information on demographics, prior diagnoses of mental health conditions, prescription fills, and depressive symptoms as measured by patient report with the patient health questionnaire \citep{kroenke_2001}. To be included in the study, patients had to be 13 years or older, had been enrolled in KPWA insurance for the past year, had a diagnosis of a depressive disorder in the 365 days before or 15 days after treatment initiation, had no antidepressant prescription fills in the prior year (excluding trazodone, doxepin, and amitriptyline, which are primarily used to treat conditions other than depression), and no diagnoses of a personality, bipolar, or psychotic disorder in the past year. We used a subset of this data to represent a pilot study to conduct sample size calculations for estimating an optimal dynamic treatment regime to minimize depression symptoms in this population.

The Patient Health Questionnaire-9 score is a measure of the severity of depressive symptoms, that is used in the KPWA health system for diagnosing depression and monitoring depressive systems \citep{kroenke_2001}. The first 8 questions yield a score that ranges from 0 to 24 (a score we shall simply refer to as the PHQ), such that higher values indicate more severe symptoms. The outcome of interest for this study was the negative of the PHQ score after 1 year of treatment to be consistent with our framework that higher values correspond with better patient outcomes. The PHQ score after 1 year is defined as the closest PHQ score recorded to exactly one year after starting treatment, but must be recorded between 305 and 425 days after beginning treatment. Initially, patients received one of 17 different antidepressants. The first stage treatment was classified as an antidepressant from the selective serotonin reuptake inhibitor (SSRI) class or an antidepressant from a different class. Section E of the online supplementary material contains a list of all the antidepressants assigned. A total of 63,060 patients received an SSRI at treatment initiation while 19,657 received an antidepressant from an alternative class. As this study follows people receiving their regular health care, patients were observed to switch to a different antidepressant or augment their initial treatment with an additional antidepressant or an antipsychotic. Our goal was to estimate an optimal dynamic treatment regime that tailors treatment based on age, gender, baseline PHQ score, and diagnosis of an anxiety disorder in the past year.

There was a significant amount of missingness and censoring in this data set. The baseline PHQ score, which is defined as a score recorded 15 days or less before treatment initiation or up to 3 days after, was observed for 34,541 (41.8\%) of the patients in the sample. Of those 34,541 patients, 8,757 (25.35\%) patients were censored due to disenrolling from the health system during the first year or administratively censored due to the study ending less than a year after beginning treatment. Of the remaining patients, 8,511 (24.6\%) had an observed PHQ at one year after treatment initiation. We also artificially censored follow-up due to patients discontinuing treatment or changing treatment more than once during the first year. Our final sample size to draw potential pilot studies from was 2,008. We constructed a pilot study by taking a simple random sample from these 2,008 remaining patients. Because of missing follow-up data and not everyone following one of the regimes of interest, it is possible our sample pilot study was not a representative sample of the population. This could lead to bias in the estimated optimal regime from the pilot. Since the purpose of this analysis is to demonstrate how to size a study we will assume our pilot is representative and not use any methods to adjust for the potential bias. Due to the potential for selection bias, the estimated regime from this pilot should not be taken to have any clinical interpretation in practice. We will consider this point further in the discussion.

Define the outcome, $Y$, as the negative PHQ score 1 year after initiating treatment. Let $A_1$ denote the first stage treatment, such that $A_1 = 1$ if prescribed an SSRI and $A_1 = 0$ if assigned a non-SSRI. Let $A_2$ denote the second stage treatment, with $A_2 = 1$ if the patient switched treatment and $A_2 = 0$ if the patient augmented treatment with an additional antidepressant or antipsychotic while staying on the initial medication. We use $\bX_1^{\beta}$ to denote a vector of age, gender, indicator variables for race, baseline PHQ score, and an indicator for the diagnosis of an anxiety disorder in the past 365 days. Let $\bX_1^{\psi}$  denote the same vector of patient characteristics with race removed as we did not consider tailoring treatment based on race. We posited the following models for the treatment-free and blip functions:
\begin{align*}
	\begin{array}{ll}
		\gamma_1(\bh_1, a_1; \psi_1) =  a_1(\psi_{1,0} + \psi_{1,1}^\T\bx_1^{\psi}), & \gamma_2(\bh_2, a_2; \psi_2) = a_2(\psi_{2,0} + \psi_{2,1}^\T\bx_1^{\psi} + \psi_{2,2}a_1 ),  \\
		g_1(\bh_1; \beta_1) = \beta_{1,0} + \beta_{1,1}^\T\bx_1^{\beta}, & g_2(\bh_2; \beta_2) = \beta_{2,0} + \beta_{2,1}^\T\bx_1^{\beta} + \beta_{2,2}a_1 .
	\end{array} 	
\end{align*}
The propensity score models, $\pi_k(\bh_k; \xi_k)$ for $k=1,2$, were estimated with logistic regression using the same set of variables as the treatment-free models.

We used a random sample of 400 patients to be included in the pilot study. We assumed that the comparison mean was given by $B_0 = -10$ as a PHQ score of greater than or equal to 10 is used to identify moderate depression. We let $\alpha = 0.05$, $\phi = 0.1$ and $\eta = 3$ so that the first condition (PWR) held if we had a 0.05 level test of $H_0: V(\bd^{\mathrm{opt}}) \leq B_0$ that had 90\% power if $V(\bd^{\mathrm{opt}}) \geq B_0 + 3$. We let $\epsilon = 1$ and $\zeta = 0.1$ so that the second condition (OPT) held if $P\{V(\widehat{\bd}_n) \geq V(\bd^{\mathrm{opt}}) - 1 \} \geq 0.9$. A confidence set for $\psi$ was constructed using the $m$-out-of-$n$ bootstrap with $\kappa = 0.2$.

Table \ref{MHRN_Psi_1_2} shows the estimated coefficients for the second and first stage blip models from the pilot data. The estimated value of the optimal regime in the pilot was given by $\widehat{V}(\widehat{\bd}_{n_0}) = -6.67$. Therefore, we found some evidence that an adaptive treatment strategy could be effective in reducing depressive symptoms, but this difference was not statistically significant in the pilot data. Note, the wide confidence intervals for the parameters of the blip models. We then applied our proposed sample size method to power the comparison of the value of the optimal regime to $B_0$ (PWR) which resulted in a sample size of  $\hat{n}(\mathcal{D}_{n_0})= 5,230$. When sizing to guarantee the value of the estimated regime is close to the value of the true optimal regime (OPT), we calculated a sample size of $\hat{n}(\mathcal{D}_{n_0})= 4,276$. Therefore, for both conditions to hold we recommend a study of size at least $\hat{n}(\mathcal{D}_{n_0})= 5,230$.

\begin{table}[h]
	\caption{
		Parameter estimates and 95\% confidence intervals for the first and second stage blip model from the KPWA pilot data}
	\begin{center}
		\begin{tabular}{| c c c |} \hline
			Covariate & Estimate & Confidence Interval \\ \hline
			\multicolumn{3}{|c|}{Second Stage Blip Model} \\
			$A_2$ & 0.06 & (-2.27, 2.39) \\
			$A_2 \times $Gender & 4.86 & (2.99, 6.73) \\
			$A_2 \times $Age & -0.04 & (-0.10, 0.02) \\
			$A_2 \times $Baseline PHQ & -0.32 & (-0.55, -0.09) \\
			$A_2 \times $Anxiety & 5.84 & (3.93,  7.75) \\
			$A_2 \times A_1$ & 5.10 & (3.11, 7.09) \\ \hline
			\multicolumn{3}{|c|}{First Stage Blip Model} \\
			$A_1$ & 2.20 & (-15.53, 11.32)\\
			$A_1 \times $Gender & 0.73 & (-3.53, 7.83)\\
			$A_1 \times $Age & -0.07 & (-0.12, 2.31)\\
			$A_1 \times $Baseline PHQ & 0.12 & (-0.39, 2.53) \\
			$A_1 \times $Anxiety & 2.19 & (-3.80, 7.67)\\ \hline
		\end{tabular}
	\end{center}
	\label{MHRN_Psi_1_2}
\end{table}

\section{Discussion}
\label{s: discussion}

We proposed a method for sizing a $K$-stage longitudinal observational study for the estimation of DTRs using a pilot study. This method is based on bootstrapping a projection interval for the value of the optimal regime. We implemented the bootstrap with oversampling to estimate the power for different sample sizes and calculate the smallest sample size that achieves the desired power. We demonstrated this method attains the desired power in finite samples in a simulation study. We also proposed a method for extending the $m$-out-of-$n$ bootstrap to multi-stage dynamic treatment regimes to obtain valid confidence intervals for the parameters indexing the treatment regime.

This paper focused on DTRs estimated via dWOLS and assumes the blip models are correctly specified. This method could be easily adapted to other regression-based estimation methods. Value-search or direct-search estimators are an alternative class of estimators for identifying optimal treatment regimes that are frequently used \citep{Orellana10, yingqi_2012, laber_2015}. We leave extensions to this class of estimators for future work. In this paper, we focused on discrete treatments with continuous outcomes. Extensions to other outcome and/or treatment types, such as survival outcomes and continuous doses, are  possible but will require careful thought.

A pilot study may include potential sources of bias, in addition to confounding, due to censoring or missing data. The resulting sample size calculations could be adjusted to account for bias by using some form of sample size inflation factor or decrease in effective sample size based on the level of censoring/missingness and how informative it is. There has also been work on using multiple imputation with bootstrapping for inference \citep{schomaker_2017, bartlett_2020} which could be adjusted to use bootstrap oversampling to conduct sample size calculations in the presence of missing data. 

Our proposed method is also reliant on having access to pilot data which are not always available. Sizing a study without pilot data would require much stronger assumptions about the underlying generative model, which we leave as future work.

\section*{Acknowledgments}

Research reported in this publication was supported by the National Institute of Mental
Health of the National Institutes of Health under Award Number R01 MH114873. The content is solely the responsibility of the authors and does not necessarily represent the official views of the National Institutes of Health. EEMM is a Canada Research Chair (Tier 1) in Statistical Methods for Precision Medicine and acknowledges the support of a chercheur de mérite senior career award from the Fonds de Recherche du Québec, Santé.

\bibliographystyle{Chicago}
\bibliography{sampleSize_dWOLS}

\end{document}